\documentstyle[prl,aps,twocolumn]{revtex}
 
\newcommand{\be}{\begin{equation}}
\newcommand{\ee}{\end{equation}}
 
\newcommand{\bea}{\begin{eqnarray}}
\newcommand{\eea}{\end{eqnarray}}
\newcommand{\ba}{\begin{array}}
\newcommand{\ea}{\end{array}}
\newcommand{\beqa}{\begin{eqnarray}}
\newcommand{\eeqa}{\end{eqnarray}}

\newcommand{\matr}{\left( \begin{array}}
\newcommand{\ematr}{\end{array} \right)}

\newcommand{\lsim}
{{\;\raise0.3ex\hbox{$<$\kern-0.75em\raise-1.1ex\hbox{$\sim$}}\;}}
\newcommand{\gsim}
{{\;\raise0.3ex\hbox{$>$\kern-0.75em\raise-1.1ex\hbox{$\sim$}}\;}}

\begin{document}
\draft
\title{Testing neutrino instability with active galactic nuclei }
\author{P. Ker\"anen \thanks{Petteri.Keranen@Helsinki.fi}, J. Maalampi \thanks{Jukka.Maalampi@Helsinki.fi}, and J.T. Peltoniemi 
\thanks{Juha.Peltoniemi@Helsinki.fi} }
\address{ Department of Physics, Box 9, 
FIN-00014 University of Helsinki, Finland.}
\date{\today}
\maketitle
\begin{abstract} 
Active galactic nuclei and gamma ray bursts at cosmological distances are sources of high-energy electron and muon neutrinos and
provide a unique  test bench for neutrino instability. The typical lifetime-to-mass ratio  one can reach there is $\tau/m\sim 500\; {\rm Mpc}/cE_{\nu}\sim 500$~s/eV.  We study the rapid decay
channel $\nu_i\to\nu_j+\phi$, where $\phi$ is a massless  or very light scalar (possibly a Goldstone boson),
and point out that one can test the coupling strength of $g_{ij}\nu_i\nu_j$ down to $g_{ij}\lsim 10^{-8}\;{\rm eV}/{m}$
by measuring the relative fluxes of $\nu_{e}$, $\nu_{\mu}$ and $\nu_{\tau}$. This is  orders of magnitude more stringent bound than what  one can obtain in other phenomena, e.g. in neutrinoless double beta decay with scalar emission. 
\end{abstract}
\pacs{13.35.Hb, 14.60.St, 95.55.Vj, 95.85.Ry}

By measuring the  spectra of the neutrinos produced in Earth's atmosphere by cosmic rays the Super-Kamiokande 
experiment has recently got evidence that the muon neutrino has a mass and that it mixes with another neutrino flavour~\cite{SuperK}. The observed deficit and angular dependence of the atmospheric  $\nu_{\mu}$  flux can be explained in terms of $\nu_{\mu}- \nu_{\tau}$ or $\nu_{\mu}-\nu_{s}$ oscillations if the squared mass difference of the oscillating states is $\Delta m^2= 3\cdot 10^{-4}-8\cdot 10^{-3} {\rm eV}^2$ and their mixing is $\sin^22\theta\gsim 0.8$. Here $\nu_{s}$ is a sterile neutrino which lacks the standard electroweak interactions.

One may draw from the Super-Kamiokande result the conclusion that  in general neutrinos have a mass and lepton flavour numbers are not separately conserved. This would consequently imply that not all neutrinos are stable particles but may decay into lighter neutrinos and other light particles. 
 If the oscillating partner of $\nu_{\mu}$ is $\nu_{\tau}$, the corresponding mass eigenstates are (ignoring all other possible mixings) $\nu_1=\cos\theta\;\nu_{\mu}-\sin\theta\;\nu_{\tau}$ and $\nu_2=\sin\theta\;\nu_{\mu}+\cos\theta\;\nu_{\tau}$, where $32^{\circ}\lsim\theta \lsim 58^{\circ}$, and assuming $m_2>m_1$, then the radiative decay  $\nu_2\to\nu_1+\gamma$ is possible. Given the small mass difference of the neutrinos as indicated by the Super-Kamiokande result, this decay is in general very slow as it proceeds through loops and is suppressed due to GIM mechanism and helicity matching.  Assuming the Standard Model (SM) interactions, the lifetime would be $\tau(\nu_{\mu}\to\nu_{\tau}+\gamma)\simeq 10^{33}$ sec~\cite{LeSc}. If one goes beyond the SM, other decay modes with shorter lifetimes may appear~\cite{MaMuRo}. The shortest lifetimes typically arise in majoron models~\cite{Majoron}, for decays with a light neutrino and a massless scalar as decay products.

The question now arises how one could detect the possible decays of  neutrinos and thereby obtain independent new evidence of their masses and mixings and information on their interactions. In this letter we shall address this question by considering the decays

\be
\nu_2\to\nu_1+\phi,
\label{decay}\ee
where $\nu_1$ and $\nu_2$ are  light and heavy mass eigenstate neutrinos (or antineutrinos)  and 
$\phi$ is a massless or very light scalar. In the simplest majoron model $\phi$ is a
massless Goldstone boson, a majoron, associated with the spontaneous breaking of  global lepton number symmetry. In the following we will not restrict ourselves to this specific model but will consider more general forms of interactions where the coupling strength may be independent of the origin of the neutrino mass. Our main result is that
the decay (\ref{decay}) could be detected by measuring the flavour content of neutrino flux from active galactic nuclei (AGN) or gamma-ray bursts (GRB's). We will find that this will provide several orders of magnitude more sensitive probe of the coupling strength than other experiments.

Let the scalar-neutrino interaction to be of the form
\be
{\cal L}_{\rm int}= g_{ij}\overline\nu^c_{i}\nu_{j}\phi \;\; +\; h.c..
\ee
As the scalar $\phi$ should be very light, it has to be dominantly a weak isospin singlet
in order to satisfy the constraint from the LEP measurement of the invisible $Z$ width~\cite{LEP}.
If the lepton number $L$ is conserved, the scalar has $L(\phi)=-2$. Examples of this kind of models are the so-called charged majoron models~\cite{CliBur} and the model where
Dirac neutrinos couple to a scalar through their singlet right-handed components via the coupling
$\overline\nu^c_{iL}\nu_{jR}\phi$~\cite{AcPaPa}. In these models the coupling strength $g_{ij}$
is not related to neutrino masses and there are no constraints from the neutrinoless double beta decay as the lepton number is conserved. If the lepton number is broken, so that $L(\phi)=0$,  neutrinos are  in general Majorana particles allowing for neutrinoless double beta decay. This is the case in the majoron model. 

In the simplest version of the model the coupling $g$ is proportional to neutrino mass, and the ensuing decay time is much too long to be relevant for our discussion. However,
other models can be easily constructed where the interaction matrix is not proportional to the mass matrix and where shorter lifetimes are possible~\cite{GelVal}.

Taking 500 Mpc as a typical distance ($D$) of AGN or GRB's and 500 TeV as a typical neutrino energy ($E$), the typical lifetime-to-mass ratio one can reach when considering decays of AGN or GRB neutrinos is
\be
\frac{\tau}{m}\simeq \frac{D}{Ec}\sim 500\frac{{\rm s}}{{\rm eV}},
\ee
which is valid for all experimentally allowed neutrino mass values.

The width of the decay $\nu_j\to \nu^c_i\;+\;\phi$ in the rest frame of $\nu_j$ is given by
\bea
\Gamma(\nu_j\to \nu^c_i\;+\;\phi) &=& \frac{g_{ij}^2}{16\pi}\frac{(m_i+m_j)^2}{m_j^3}\delta m_{ji}^2\nonumber\\
&\simeq &\frac{g_{ij}^2}{16\pi} m_j,
\eea
where $\delta m_{ji}^2 = m_j^2-m_i^2$. The lifetime in the rest system of the observer, where $\nu_j$ has the energy $E_j$, is then
\bea
\tau (\nu_j\to \nu^c_i&\;+&\;\phi)=\frac{E_j}{m_j\Gamma}\nonumber\\
&\simeq& g_{ij}^{-2}\cdot 3.3\; {\rm sec}\cdot\left(\frac{E_j}{100\;{\rm TeV}}\right)\left(\frac{1\;{\rm eV}}{m_{j}}\right)^2.
\eea
For a relativistic decaying neutrino the decay distance is then given by
\be
L=c\tau\simeq  g_{ij}^{-2}\cdot 1.0\times 10^{9}\; {\rm m}\cdot\left(\frac{E_j}{100\;{\rm TeV}}\right)\left(\frac{1\;{\rm eV}}{m_{j}}\right)^2.
\label{length}
\ee

In order to see the effect of the decay, one should observe a sufficiently abundant neutrino beam from a flight distance of $D\gsim L$. As the numbers in (\ref{length}) indicate, astrophysical or cosmological distances are preferred, and AGN and GRB's stand out as possible beam sources. Note, however, that neutrino decays may be relevant also  with respect to the atmospheric neutrino problem where smaller distances and energies are involved~\cite{BaLePaWe}. The highest energy cosmic rays are known to originate in AGN and GRB's, and they are assumed to produce also neutrinos from the decay of charged pions through the chain $\pi^+\to \nu_{\mu}\mu^+\to \nu_{\mu}e^+\nu_e\bar\nu_{\mu}$~\cite{AGN}. The primary process where the pions are produced is the interactions of accelerated protons with the photons via $p\gamma\to n\pi^+$ in the source near the $\Delta$ resonance. 
The energies of neutrinos originating in AGN jets are typically $10^{5}$ TeV, but less energetic neutrinos are abundantly produced in accretion disks~\cite{disk}. The typical energies of GRB neutrinos are around 100 TeV. The most powerful AGN and GRB's are located at distances of 100-1000 Mpc or more. AGN and GRB neutrinos whose energy exceeds 10 TeV can be distinguished from the atmospheric neutrino background~\cite{Gandhi}.

If we take $E=100$ TeV, $D=500$ Mpc as face values, we will find from (\ref{length})
that the values of the coupling constant down to

\be
g_{ij}\simeq 10^{-8}\; \frac{\rm eV}{m}
\ee
  could in principle be probed by observing the neutrino flux from AGN and GRB's and measuring the relative fractions of its flavour components. 

The neutrino beam created at AGN and GRB's consists of $\nu_e$ and $\nu_{\mu}$ with no
substantial $\nu_{\tau}$ component. If we assume that the oscillating partner of atmospheric $\nu_{\mu}$ is the tau neutrino, oscillations will create a $\nu_{\tau}$ component into the AGN and GRB neutrino beam of about the same size as the $\nu_{\mu}$ component as the mixing is almost
maximal. Actually, with such a baseline the presence of a $\nu_{\tau}$ component
would indicate neutrino oscillations with $\Delta m^2$ as low as $10^{-17}$ eV$^2$~\cite{LePa,HalSal}. 

In the case of almost maximal mixing the decay will not dramatically change the  content of the beam from the fifty-fifty division as both decaying and produced mass eigenstate neutrinos  contain  $\nu_{\mu}$ and $\nu_{\tau}$ components in almost equal fractions. At most, the  abundance ratio of $\nu_{\tau}/\nu_{\mu}$ in the beam is changed due to the decay by the factor $\cot\theta\simeq 1.6$, corresponding to the lowest mixing angle allowed by the Super-Kamiokande measurement. 

In contrast,  a very clear signal would be available in the case where the decay happens into a third neutrino mass eigenstate consisting mostly of the electron neutrino $\nu_e$ but having no substantial $\nu_{\mu}$ and $\nu_{\tau}$ components. Actually, in the case where the $\nu_{\mu}-\nu_{\tau}$ mixing is responsible for the atmospheric neutrino signal, the electron neutrino should mix with a sterile neutrino in order to explain the solar neutrino deficit problem, i.e. there should exist another two mass eigenstates
$\nu_3=\cos\phi\;\nu_{e}-\sin\phi\;\nu_{s}$ and $\nu_4=\sin\phi\;\nu_{e}+\cos\phi\;\nu_{s}$, where $\nu_s$
is a sterile neutrino. Three solutions are still possible; two of them are based on the MSW mechanism~\cite{MSW}, one with  small mixing angle ($\sin^22\theta= 5.5\times 10^{-3}$), one with large mixing angle ($\sin^22\theta\simeq 0.8$), and the third one is based on vacuum oscillations~\cite{Vacuum} with large mixing angle ($\sin^22\theta\simeq 0.75$). For the MSW solutions the squared mass difference
$\Delta m_{34}^2$ is  $5\times 10^{-6}$ eV$^2$ and for the vacuum oscillation solution $10^{-10}$ eV$^2$. The most natural mass hierarchy would be that $m_1\sim m_2\gg m_3\sim m_4$. Because of the strong mixing of $\nu_{\mu}$ and $\nu_{\tau}$,  both heavy components of the original AGN neutrino beam,
$\nu_{1}$ and $\nu_{2}$, would in this case be expected to decay into the lighter mass states $\nu_{3}$ and $\nu_{4}$, resulting in the disappearance of the expected AGN muon neutrinos, as well as the non-appearance
of tau neutrinos in spite of the $\nu_{\mu}-\nu_{\tau}$ mixing.

If the atmospheric neutrino behaviour is due to mixing between $\nu_{\mu}$ and a sterile neutrino $\nu_{s}$, then the solar neutrino deficit problem could be solved through a $\nu_{e}-\nu_{\tau}$ mixing. This case differs from the previous one in that now the disappearance of  muon neutrinos could be accompanied with the appearance of tau neutrinos as a result of decays. If  the $\nu_{e}-\nu_{\tau}$ mixing is large, as is the case when the solar neutrino deficit results from vacuum oscillations or a large-mixing MSW effect, then  a large 
$\nu_{\tau}$  component would be generated also through $\nu_{e}-\nu_{\tau}$ oscillations, but in the case of the small-mixing MSW effect $\nu_{\tau}$'s would appear almost exclusively due to decays.

To experimentally observe the possible decays of AGN and GRB neutrinos it is hence essential to
measure the relative fluxes of $\nu_{e}$, $\nu_{\mu}$ and $\nu_{\tau}$. While the muon neutrino will be relatively easy to detect by tracking the muon produced in charged-current interaction in the vicinity of a Cherenkov detector in deep water or ice, the detection of electron and in particularly tau neutrinos is less straightforward.
The electron neutrinos create an electromagnetic cascade which can be detected by optical or radio technique
~\cite{BaHaPr}. For the detection of the tau neutrino two methods has been recently proposed:   by double-bang events from the production and decay of the $\tau$ lepton~\cite{LePa}, and by the absence of absorption by the Earth~\cite{HalSal}. 
In general, the effective elimination of the background  requires that one should look at up-going neutrinos. 

Depending on the scenario realized in Nature, one can get information on various coupling constants
$g_{ij}$ ($i,j=1,2,3,0$) or their combinations by measuring the AGN and GRB neutrino flux in the Earth. As mentioned already, the sensitivity is in principle on the level of $g_{ij}\lsim 10^{-8}$.
Limits obtained from other astropysical objects, such as supernovas, and from early Universe are in general less stringent (see Ref.~\cite{Raffelt} for a review).  The strongest laboratory bounds on the scalar couplings $g_{ij}\nu_i\nu_j$
are due to the scalar-emitting neutrinoless double beta decay~\cite{betabeta}. They apply to the couplings where both the mass eigenstates $\nu_i$ and $\nu_j$ contain $\nu_e$ flavour component, and the magnitude of the upper bound for $g_{ij}$ depends on whether the scalar is a Goldstone boson or not  as this affects the form of the electron spectrum. In the case $\phi$ is not a Goldstone boson the constraint is $g_{ij}\lsim 10^{-4}$ and in the case it is $g_{ij}\lsim 0.1$. These limits do not compete with the limits one can achieve from AGN and GRB's.
Let us note that exotic neutrino interactions of the AGN and GRB neutrinos can be also probed via their interactions with the cosmic neutrino background~\cite{Keranen}.

To summarize, we have pointed out that if the neutrino beam from  AGN and GRB's contains neither $\nu_{\mu}$'s nor $\nu_{\tau}$'s  the only explanation in the framework we have considered  would be the decay of the $\nu_{\mu}$ containing mass eigenstates into a lighter neutrino and a scalar.
If one observes a $\nu_{\tau}$ component in the neutrino flux but no $\nu_{\mu}$ component
this will also indicate neutrino decay. If there is also a $\nu_{\mu}$ component in the beam, then the $\nu_{\tau}$ component could also result from oscillations. Because of large distance to the source this test of neutrino stability is  orders of magnitude more sensitive than other astrophysical and cosmological, as well as laboratory, tests.
\\

We are indebted to Matts Roos for useful discussions. One of us (P.K.) expresses his gratitude to the Magnus Ehrnrooth Foundation for a grant. This work has been supported by the Academy of Finland under the project no. 40677.

\end{document}